\documentclass[twoside]{article}
\usepackage{fleqn,espcrc2,amssymb}

\usepackage{graphicx}

\newcommand{\AmS}{{\protect\the\textfont2
  A\kern-.1667em\lower.5ex\hbox{M}\kern-.125emS}}
\title{$ab$-plane resistivity and possible charge stripe ordering in strongly underdoped
La$_{2-x}$Sr$_{x}$CuO$_{4}$ single crystals}

\author{\mbox{R.S. Gonnelli\address{INFM - Dipartimento di Fisica, Politecnico di Torino, c.so Duca degli
Abruzzi 24, 10129 Torino, Italy},
         V.A. Stepanov\address{P.N. Lebedev Physical Institute, Russian Academy of Sciences,
         SU-117924 Moscow, Russia},
         A. Morello$^{\rm a}$,
         G.A. Ummarino$^{\rm a}$,
         D. Daghero$^{\rm a}$,
         F. Licci\address{Istituto MASPEC del CNR, Parco Area delle Scienze, 37/A - 43010 Loc. Fontanini -  Parma, Italy}
         and
         G. Ubertalli\address{Dipartimento di Scienza dei Materiali e Ing. Chimica, Politecnico di Torino, 10129 Torino, Italy}}}

\begin{document}

\begin{abstract}
We have measured the $ab$-plane resistivity of
La$_{2-x}$Sr$_x$CuO$_4$ single crystals with small Sr content
(x=0.052$\div$0.075) between 4.2 and $300\,$K by using the AC Van
der Pauw technique. As recently suggested by Ichikawa {\it et
al}., the deviation from the linearity of the
$\rho_{\mathrm{ab}}(T)$ curve starting at a temperature
T$_{\mathrm{ch}}$ can be interpreted as due to a progressive
slowing down of the fluctuations of pre-formed charge stripes. An
electronic transition of the stripes to a more ordered phase could
instead be responsible for some very sharp anomalies present in
the $\rho_{\mathrm{ab}}(T)$  of superconducting samples just above
$T_{\mathrm{c}}$. \vspace{-4mm}
\end{abstract}

\maketitle
%
There is a growing experimental evidence \cite{Tranquada96,Imai99}
for the existence of charge stripes in La$_{2-x}$Sr$_{x}$CuO$_{4}$
(LSCO) and the related compounds obtained by substitutions.
Successive ordering transitions of these stripes, which seem to
have a metallic character \cite{Tranquada99a}, possibly occur when
the temperature is lowered. One would expect that, in
superconducting samples, the phase transitions of the stripes
occurring above $T_{\mathrm{c}}$ leave a mark on the electrical
resistivity \cite{Tranquada99a}. In the present work, we study the
in-plane resistivity $\rho_\mathrm{ab}(T)$ of strongly underdoped
single crystals of LSCO, which exhibits, besides the usual
features previously observed, an anomalous peak just above
$T_{\mathrm{c}}$. We relate it 
to an electronic transition of the stripes to a more ordered phase
\cite{Emery}.

\begin{figure}[t]
\vspace{-9mm}
\includegraphics[keepaspectratio,width=7.5cm]{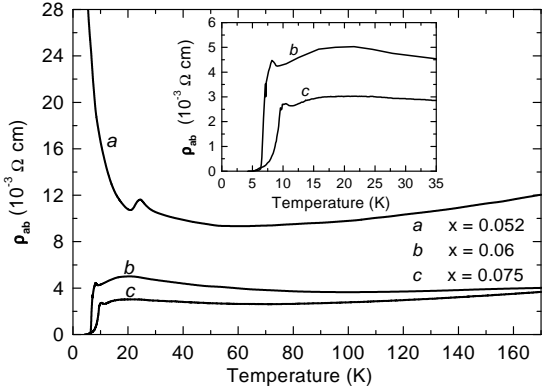}
\vspace{-13mm} \caption{\small{The $ab$-plane resistivity of our
LSCO samples with $x = 0.052, 0.06$ and $0.075$ (curves {\it a},
{\it b} and {\it c}, respectively). The inset shows an enlargement
of the low-temperature part of curves {\it b} and {\it c}.}}
\vspace{-10mm}\label{fig1}
\end{figure}

We carefully investigated the temperature dependence of the
$ab$-plane resistivity  of 
LSCO single crystals with small Sr doping (\mbox{$0.052\leq x\leq
0.075$}). The crystals, having typical dimensions of about
$1\times1\times0.3\,$mm$^3$, were grown by slowly cooling a
non-stoichiometric melt. The chemical composition of every crystal
was determined by means of EDS microprobe analysis and only
crystals having a homogeneous Sr content were selected. In order
to increase the precision of the $ab$-plane measure, we used an AC
version of the standard four-probes eight-measures Van der Pauw
method injecting in the crystals AC currents of $100\div
300\,\mu$A at $133\,$Hz. Details on the experimental technique are
given elsewhere \cite{Gonn99}. Fig.~\ref{fig1} shows the results.
The crystals with $x=0.052$ are not superconducting and show a
low-temperature semiconducting-like behaviour of the resistivity,
apart from an anomalous local maximum at about $25\,$K (curve
\emph{a} in Fig.~\ref{fig1}). The crystals with $x = 0.06$ and
0.075 show well-defined superconducting transitions with midpoints
at $T_\mathrm{c} = 6.7\,$K ($\Delta T_\mathrm{c} = 0.6\,$K) and
$T_\mathrm{c} = 8.9\,$K ($\Delta T_\mathrm{c} = 2.2\,$K),
respectively (curves \emph{b} and \emph{c}). However, very sharp
anomalous peaks are present at temperatures about $2\,$K greater
than $T_\mathrm{c}$. The inset of Fig.~\ref{fig1} shows these
peaks in greater detail. No anomalies are present in the AC
susceptibility of the same crystals.
%

We first discuss the general behaviour of the resistivity curves.
The $\rho_\mathrm{ab}(T)$ at \mbox{$150 < T < 300\,$K} is almost
perfectly linear. We then fitted the resistivity in this region by
using a linear model: \mbox{$\rho_\mathrm{ab} (T) = \alpha T +
\beta$}. Following a recent paper of Ichikawa {\it et al.}
\cite{Tranquada99a}, we calculated the normalized resistivity
\mbox{$\rho_\mathrm{ab} (T) / (\alpha T + \beta)$}. The results
for the superconducting samples are shown in Fig.~2. In Ref.~3 the
deviations of the $\rho_\mathrm{ab}(T)$ from the linearity
exceeding a certain threshold (dependent on $x$) have been
associated to the localization of pre-formed stripes, which occurs
at a temperature $T_\mathrm{ch}$ independently determined by Cu
NQR measurements \cite{Imai99}. This correlation has been observed
in Nd-substituted crystals but also in crystals without Nd, having
$x$=0.10 and $x$=0.12 \cite{Tranquada99a}. Following the same
approach, we obtained the stripe localization temperatures
$T_\mathrm{ch}$ in our samples by crossing the normalized
resistivities with appropriate thresholds extrapolated from the
data of Ref.~3. Solid circles on curves \emph{a} and \emph{b} of
Fig.~2 show the results of this operation. The inset of Fig.~2
shows the temperatures $T_\mathrm{ch}$ of our crystals versus the
Sr-doping (solid squares): they are in {\it excellent agreement}
with $T_\mathrm{ch}$ values determined by means of NQR measurement
in LSCO crystals with different doping \cite{Imai99} (open
squares). A further decrease of the temperature below
$T_\mathrm{ch}$ produces a continuous slowing-down of the stripe
motion: stripes start to be pinned (by defects or by intrinsic low
doping) and $\rho_\mathrm{ab} (T)$ shows an upturn up to a peak
followed by the region dominated by superconducting fluctuations.

One could argue that the small peaks at $T_\mathrm{o}$ about 2~K
above $T_\mathrm{c}$ (see the inset of Fig.~2) may be due to
measurement errors, $\rho_{\mathrm{c}}$ interferences in
$\rho_{\mathrm{ab}}$, or crystal inhomogeneities (i.e. parts of
the crystal are not superconducting and contribute to
$\rho_{\mathrm{ab}}$ with a semiconducting-like term just above
$T_\mathrm{c}$).
We can reply to these objections that (i) the Van der Pauw's
four-probes AC technique provides four resistance measures and all
of them show very reproducible peaks at $T_\mathrm{o}$; (ii) a
detailed data analysis suggests that the resistivity contribution
of \emph{c}-axis terms or different-doping islands is unable to
give peaks as those observed \cite{Gonn99}. Another possible
explanation for their origin could be a \emph{structural}
low-temperature transition, similar to that observed at higher
temperature in Nd-substituted samples \cite{Tranquada99a}, but
there are no other evidences of such a transition near
$T_{\mathrm{c}}$ in LSCO.

We speculate that an \emph{electronic} phase transition  from a
\emph{nematic} stripe phase to a more ordered \emph{smectic} one
(``stripe glass'') \cite{Emery} is responsible for the anomalies
near $T_{\mathrm{c}}$. This scenario is supported by theoretical
arguments \cite{Emery} and experimental analogies with the
behaviour of some metal dichalcogenides at the charge-density wave
transition and it is consistent with the coexistence of cluster
spin-glass and superconductivity recently observed below $\sim$
5~K in LSCO with $x=0.06$ \cite{Julien}.
%
\begin{figure}[t]
\vspace{-10mm}
\includegraphics[keepaspectratio,width=7.5cm]{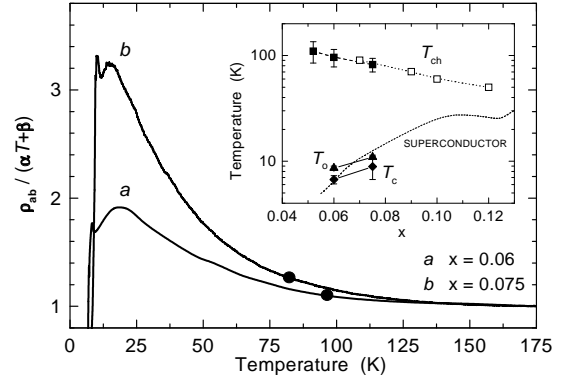}
\vspace{-13mm} \caption{\small{The normalized $ab$-plane
resistivity of our samples with $x = 0.06$ (curve {\it a}) and
$0.075$ (curve {\it b}). The inset shows the temperatures
$T_\mathrm{ch}$, $T_\mathrm{o}$ and $T_\mathrm{c}$ determined from
our resistivity data (solid symbols) and $T_\mathrm{ch}$ from NQR
data \cite{Imai99} (open squares) as function of the doping (for
details see the text).}} \vspace{-8mm}\label{fig2}
\end{figure}

\vspace{-2mm}
\begin{thebibliography}{1}
{\small
\bibitem{Tranquada96} J.M.~Tranquada {\it et al}., 
Phys.~Rev.~B 54 (1996) 7489.
\bibitem{Imai99} A.W.~Hunt {\it et al}., 
Phys.~Rev.~Lett. 82 (1999) 4300; P.M.~Singer {\it et al}., 
cond-mat/9906140. 
\bibitem{Tranquada99a} N.~Ichikawa {\it et al}., 
cond-mat/9910037; J.M.~Tranquada {\it et al}., 
Phys.~Rev.~B 59 (1999) 14712. 
\bibitem{Emery} \mbox{S.A.$\,$Kivelson$\,$and$\,$V.J.$\,$Emery,$\,$cond-mat/9809082}; S.A.~Kivelson {\it et al}.,
Nature 393 (1998) 550. 
\bibitem{Gonn99} R.S. Gonnelli {\it et al}., to be published.
\bibitem{Julien} \mbox{M.-H.~Julien {\it et al}.,$\,$Phys.$\,$Rev.$\,$Lett.$\,$83$\,$(1999)$\,$604.}}
\end{thebibliography}
\end{document}